\documentclass[conference]{IEEEtran}

\IEEEoverridecommandlockouts

\usepackage{amsmath,amsthm,relsize}
\usepackage{amsfonts, amssymb, cuted}
\usepackage{cleveref}
\usepackage{mathtools}
\usepackage{algorithm}
\usepackage[noend]{algpseudocode}
\usepackage{cite}
\usepackage{xcolor}
\usepackage{soul}
\usepackage{graphicx}
\usepackage{tikz}
\usepackage{pgfplotstable}
\usepackage{pgfplots}
\usepackage{nicefrac}
\usepackage{enumitem}
\usepackage{support-caption}
\usepackage{subcaption}
\usepackage[font=footnotesize]{caption}
\usepackage{multirow}
\pgfplotsset{compat=1.15}
\usepackage{relsize}
\usetikzlibrary{patterns}
\usepackage{acro}

\newcommand{\vbar}{\raisebox{.17ex}{\rule{.04em}{1.35ex}}}
\newcommand{\vbarind}{\raisebox{.01ex}{\rule{.04em}{1.1ex}}}
\newcommand{\R}{\ifmmode{\rm I}\hspace{-.2em}{\rm R} \else ${\rm I}\hspace{-.2em}{\rm R}$ \fi}
\newcommand{\T}{\ifmmode{\rm I}\hspace{-.2em}{\rm T} \else ${\rm I}\hspace{-.2em}{\rm T}$ \fi}
\newcommand{\N}{\ifmmode{\rm I}\hspace{-.2em}{\rm N} \else \mbox{${\rm I}\hspace{-.2em}{\rm N}$} \fi}
\newcommand{\B}{\ifmmode{\rm I}\hspace{-.2em}{\rm B} \else \mbox{${\rm I}\hspace{-.2em}{\rm B}$} \fi}
\newcommand{\Hil}{\ifmmode{\rm I}\hspace{-.2em}{\rm H} \else \mbox{${\rm I}\hspace{-.2em}{\rm H}$} \fi}
\newcommand{\C}{\ifmmode\hspace{.2em}\vbar\hspace{-.31em}{\rm C} \else \mbox{$\hspace{.2em}\vbar\hspace{-.31em}{\rm C}$} \fi}
\newcommand{\Cind}{\ifmmode\hspace{.2em}\vbarind\hspace{-.25em}{\rm C} \else \mbox{$\hspace{.2em}\vbarind\hspace{-.25em}{\rm C}$} \fi}
\newcommand{\Q}{\ifmmode\hspace{.2em}\vbar\hspace{-.31em}{\rm Q} \else \mbox{$\hspace{.2em}\vbar\hspace{-.31em}{\rm Q}$} \fi}
\newcommand{\Z}{\ifmmode{\rm Z}\hspace{-.28em}{\rm Z} \else ${\rm Z}\hspace{-.28em}{\rm Z}$ \fi}

\DeclareAcronym{AWGN}{
    short = AWGN,
    long = additive white Gaussian noise,
    list = Additive White Gaussian Noise,
    tag = abbrev
}

\DeclareAcronym{ADMM}{
    short = ADMM,
    long = alternating direction method of multipliers,
    list = Alternating Direction Method of Multipliers,
    tag = abbrev
}

\DeclareAcronym{MGMC}{
    short = MGMC,
    long = multi-group multi-casting,
    list = multi-group multi-casting,
    tag = abbrev
}

\DeclareAcronym{SGMC}{
    short = SGMC,
    long = single-group multi-casting,
    list = single-group multi-casting,
    tag = abbrev
}
\DeclareAcronym{AoA}{
    short = AoA,
    long = angle-of-arrival,
    list = Angle-of-Arrival,
    tag = abbrev
}

\DeclareAcronym{AoD}{
    short = AoD,
    long = angle-of-departure,
    list = Angle-of-Departure,
    tag = abbrev
}

\DeclareAcronym{KKT}{
    short = KKT,
    long = Karush-Kuhn-Tucker,
    list = Karush-Kuhn-Tucker,
    tag = abbrev
}

\DeclareAcronym{MMF}{
    short = MMF,
    long = max-min-fairness,
    list = max-min-fairness,
    tag = abbrev
}

\DeclareAcronym{WMMF}{
    short = WMMF,
    long = weighted max-min-fairness,
    list = max-min-fairness,
    tag = abbrev
}

\DeclareAcronym{BB}{
    short = BB,
    long = base band,
    list = Base Band,
    tag = abbrev
}

\DeclareAcronym{BC}{
    short = BC,
    long = broadcast channel,
    list = Broadcast Channel,
    tag = abbrev
}

\DeclareAcronym{BS}{
    short = BS,
    long = base station,
    list = Base Station,
    tag = abbrev
}

\DeclareAcronym{BR}{
    short = BR,
    long = best response,
    list = Best Response, 
    tag = abbrev
}

\DeclareAcronym{CB}{
    short = CB,
    long = coordinated beamforming,
    list = Coordinated Beamforming,
    tag = abbrev
}

\DeclareAcronym{CC}{
    short = CC,
    long = coded caching,
    list = Coded Caching,
    tag = abbrev
}

\DeclareAcronym{CE}{
    short = CE,
    long = channel estimation,
    list = Channel Estimation,
    tag = abbrev
}

\DeclareAcronym{CoMP}{
    short = CoMP,
    long = coordinated multi-point transmission,
    list = Coordinated Multi-Point Transmission,
    tag = abbrev
}

\DeclareAcronym{CRAN}{
    short = C-RAN,
    long = cloud radio access network,
    list = Cloud Radio Access Network,
    tag = abbrev
}

\DeclareAcronym{CSE}{
    short = CSE,
    long = channel specific estimation,
    list = Channel Specific Estimation,
    tag = abbrev
}

\DeclareAcronym{CSI}{
    short = CSI,
    long = channel state information,
    list = Channel State Information,
    tag = abbrev
}

\DeclareAcronym{CSIT}{
    short = CSIT,
    long = channel state information at the transmitter,
    list = Channel State Information at the Transmitter,
    tag = abbrev
}

\DeclareAcronym{CU}{
    short = CU,
    long = central unit,
    list = Central Unit,
    tag = abbrev
}

\DeclareAcronym{D2D}{
    short = D2D,
    long = device-to-device,
    list = Device-to-Device,
    tag = abbrev
}

\DeclareAcronym{DE-ADMM}{
    short = DE-ADMM,
    long = direct estimation with alternating direction method of multipliers,
    list = Direct Estimation with Alternating Direction Method of Multipliers,
    tag = abbrev
}

\DeclareAcronym{DE-BR}{
    short = DE-BR,
    long = direct estimation with best response,
    list = Direct Estimation with Best Response,
    tag = abbrev
}

\DeclareAcronym{DE-SG}{
    short = DE-SG,
    long = direct estimation with stochastic gradient,
    list = Direct Estimation with Stochastic Gradient,
    tag = abbrev
}

\DeclareAcronym{DFT}{
	short = DFT,
	long = discrete fourier transform,
	list = Discrete Fourier Transform,
	tag = abbrev
}

\DeclareAcronym{DoF}{
    short = DoF,
    long = degrees of freedom,
    list = Degrees of Freedom,
    tag = abbrev
}

\DeclareAcronym{DL}{
    short = DL,
    long = downlink,
    list = Downlink,
    tag = abbrev
}

\DeclareAcronym{GD}{
	short = GD, 
	long = gradient descent,
	list = Gradeitn Descent,
	tag = abbrev
}

\DeclareAcronym{IBC}{
    short = IBC,
    long = interfering broadcast channel,
    list = Interfering Broadcast Channel,
    tag = abbrev
}

\DeclareAcronym{i.i.d.}{
    short = i.i.d.,
    long = independent and identically distributed,
    list = Independent and Identically Distributed,
    tag = abbrev
}

\DeclareAcronym{JP}{
    short = JP,
    long = joint processing,
    list = Joint Processing,
    tag = abbrev
}

\DeclareAcronym{LOS}{
	short = LOS,
	long = line-of-sight,
	list = Line-of-Sight,
	tag = abbrev
}

\DeclareAcronym{LS}{
    short = LS,
    long = least squares,
    list = Least Squares,
    tag = abbrev
}

\DeclareAcronym{LTE}{
    short = LTE,
    long = Long Term Evolution,
    tag = abbrev
}

\DeclareAcronym{LTE-A}{
    short = LTE-A,
    long = Long Term Evolution Advanced,
    tag = abbrev
}

\DeclareAcronym{MIMO}{
    short = MIMO,
    long = multiple-input multiple-output,
    list = Multiple-Input Multiple-Output,
    tag = abbrev
}

\DeclareAcronym{MISO}{
    short = MISO,
    long = multiple-input single-output,
    list = Multiple-Input Single-Output,
    tag = abbrev
}

\DeclareAcronym{MAC}{
    short = MAC,
    long = multiple access channel,
    list = Multiple Access Channel,
    tag = abbrev
}

\DeclareAcronym{MSE}{
    short = MSE,
    long = mean-squared error,
    list = Mean-Squared Error,
    tag = abbrev
}

\DeclareAcronym{MMSE}{
    short = MMSE,
    long = minimum mean-squared error,
    list = Minimum Mean-Squared Error,
    tag = abbrev
}

\DeclareAcronym{mmWave}{
	short = mmWave,
	long = millimeter wave,
	list = Millimeter Wave,
	tag = abbrev
}

\DeclareAcronym{MU-MIMO}{
    short = MU-MIMO,
    long = multi-user \ac{MIMO},
    list = Multi-User \ac{MIMO},
    tag = abbrev
}

\DeclareAcronym{OTA}{
    short = OTA,
    long = over-the-air,
    list = Over-the-Air,
    tag = abbrev
}

\DeclareAcronym{PSD}{
    short = PSD,
    long = positive semidefinite,
    list = Positive Semidefinite,
    tag = abbrev
}

\DeclareAcronym{QoS}{
	short = QoS,
	long = quality of service,
	list = Quality of Service,
	tag = abbrev
}

\DeclareAcronym{RCP}{
	short = RCP,
	long = remote central processor,
	list = Remote Central Processor,
	tag = abbrev
}

\DeclareAcronym{RRH}{
    short = RRH,
    long = remote radio head,
    list = Remote Radio Head,
    tag = abbrev
}

\DeclareAcronym{RSSI}{
    short = RSSI,
    long = received signal strength indicator,
    list = Received Signal Strength Indicator,
    tag = abbrev
}

\DeclareAcronym{RX}{
	short = RX,
	long = receiver,
	list = Receiver,
	tag = abbrev
}

\DeclareAcronym{SCA}{
    short = SCA,
    long = successive-convex-approximation,
    list = Successive-Convex-Approximation,
    tag = abbrev
}

\DeclareAcronym{SG}{
    short = SG,
    long = stochastic gradient,
    list = Stochastic Gradient,
    tag = abbrev
}

\DeclareAcronym{SIC}{
    short = SIC,
    long = successive interference cancellation,
    list = Successive Interference Cancellation,
    tag = abbrev
}

\DeclareAcronym{SNR}{
    short = SNR,
    long = signal-to-noise-ratio,
    list = Signal-to-Noise Ratio,
    tag = abbrev
}

\DeclareAcronym{SDR}{
    short = SDR,
    long = semi-definite-relaxation,
    list = semi-definite-relaxation,
    tag = abbrev
}

\DeclareAcronym{SINR}{
    short = SINR,
    long = signal-to-interference-plus-noise ratio,
    list = Signal-to-Interference-plus-Noise Ratio,
    tag = abbrev
}

\DeclareAcronym{SOCP}{
	short = SOCP, 
	long = second order cone program,
	list = Second Order Cone Program,
	tag = abbrev
}

\DeclareAcronym{SSE}{
    short = SSE,
    long = stream specific estimation,
    list = Stream Specific Estimation,
    tag = abbrev
}

\DeclareAcronym{SVD}{
	short = SVD,
	long = singular value decomposition,
	list = Singular Value Decomposition,
	tag = abbrev
}

\DeclareAcronym{TDD}{
	short = TDD,
	long = time division duplex,
	list = Time Division Duplex,
	tag = abbrev
}

\DeclareAcronym{TX}{
	short = TX,
	long = transmitter,
	list = Transmitter,
	tag = abbrev
}

\DeclareAcronym{UE}{
    short = UE,
    long = user equipment,
    list = User Equipment,
    tag = abbrev
}

\DeclareAcronym{UL}{
    short = UL,
    long = uplink,
    list = Uplink,
    tag = abbrev
}

\DeclareAcronym{ULA}{
	short = ULA,
	long = uniform linear array,
	list = Uniform Linear Array,
	tag = abbrev
}

\DeclareAcronym{UPA}{
    short = UPA,
    long = uniform planar array,
    list = Uniform Planar Array,
    tag = abbrev
}

\DeclareAcronym{WMMSE}{
    short = WMMSE,
    long = weighted minimum mean-squared error,
    list = Weighted Minimum Mean-Squared Error,
    tag = abbrev
}

\DeclareAcronym{WMSEMin}{
    short = WMSEMin,
    long = weighted sum \ac{MSE} minimization,
    list = Weighted sum \ac{MSE} Minimization,
    tag = abbrev
}

\DeclareAcronym{WBAN}{
	short = WBAN,
	long = wireless body area network,
	list = Wireless Body Area Network,
	tag = abbrev
}

\DeclareAcronym{WSRMax}{
    short = WSRMax,
    long = weighted sum rate maximization,
    list = Weighted Sum Rate Maximization,
    tag = abbrev
}

\newtheorem{exmp}{Example}
\theoremstyle{definition}

\newcommand{\CA}[0]{{\mathcal{A}}}
\newcommand{\CB}[0]{{\mathcal{B}}}

\newcommand{\CD}[0]{{\mathcal{D}}}
\newcommand{\CE}[0]{{\mathcal{E}}}
\newcommand{\CF}[0]{{\mathcal{F}}}

\newcommand{\CR}[0]{{\mathcal{R}}}

\newcommand{\Ba}[0]{{\mathbf{a}}}

\newcommand{\Bk}[0]{{\mathbf{k}}}

\newcommand{\Bp}[0]{{\mathbf{p}}}

\newcommand{\Bw}[0]{{\mathbf{w}}}
\newcommand{\Bx}[0]{{\mathbf{x}}}

\newcommand{\BA}[0]{{\mathbf{A}}}

\newcommand{\BV}[0]{{\mathbf{V}}}

\newcommand{\Brk}[0]{{\Bar{k}}}

\newcommand{\Brn}[0]{{\Bar{n}}}

\newcommand{\Brq}[0]{{\Bar{q}}}

\newcommand{\Brt}[0]{{\Bar{t}}}

\newcommand{\BrW}[0]{{\Bar{W}}}

\newcommand{\FillGray}[3]{\filldraw[gray!50](#3-1+0.1,#1-#2+0.1) rectangle (#3-0.1,#1-#2+1-0.1)}
\newcommand{\FillBlack}[3]{\filldraw[black!70](#3-1+0.1,#1-#2+0.1) rectangle (#3-0.1,#1-#2+1-0.1)}
\newcommand{\FillHatch}[3]{\fill[pattern=crosshatch, pattern color=black!65](#3-1,#1-#2)rectangle(#3,#1-#2+1)}
\newcommand{\PutText}[4]{\node[] at (#3-1+0.5,#1-#2+0.5) {\small #4}}

\newcommand{\subparagraph}{}
\usepackage{titlesec}
\titlespacing\section{3pt}{6pt plus 4pt minus 2pt}{6pt plus 2pt minus 2pt}
\titlespacing\subsection{3pt}{4pt plus 4pt minus 2pt}{4pt plus 2pt minus 2pt}
\titlespacing\subsubsection{3pt}{3pt plus 4pt minus 2pt}{0pt plus 2pt minus 3pt}

\title{Low-Subpacketization Multi-Antenna\\Coded Caching for Dynamic Networks}

\begin{document}

\author{\IEEEauthorblockN{MohammadJavad Salehi\IEEEauthorrefmark{1}, Emanuele Parinello\IEEEauthorrefmark{2}, Hamidreza Bakhshzad Mahmoodi\IEEEauthorrefmark{1}, and Antti T\"olli\IEEEauthorrefmark{1}} \\
\IEEEauthorblockA{
    \IEEEauthorrefmark{1}Centre for Wireless Communications, University of Oulu, 90570 Oulu, Finland \\
    \IEEEauthorrefmark{2}Communication Systems Department, Eurecom, Sophia Antipolis, 06410 Biot, France\\
    \textrm{E-mail: \{firstname.lastname\}@oulu.fi, \{firstname.lastname\}@eurecom.fr}
    }
\thanks{
This work is supported by the Academy of Finland under grant no. 318927 (6Genesis Flagship), and by Vaikuttavuuss\"a\"ati\"o under the project Directional Data Delivery for Wireless Immersive Digital Environments (3D-WIDE).}\vspace{-10pt}
}

\maketitle

\begin{abstract}
Multi-antenna coded caching combines a global caching gain, proportional to the total cache size in the network, with an additional spatial multiplexing gain that stems from multiple transmitting antennas. However, classic centralized coded caching schemes are not suitable for dynamic networks as they require prior knowledge of the number of users to indicate what data should be cached at each user during the placement phase. On the other hand, fully decentralized schemes provide comparable gains to their centralized counterparts only when the number of users is very large. In this paper, we propose a novel multi-antenna coded caching scheme for dynamic networks, where instead of defining individual cache contents, we associate users with a limited set of predefined caching profiles. Then, during the delivery phase, we aim at achieving a combined caching and spatial multiplexing gain, comparable to a large extent with the ideal case of fully centralized schemes. The resulting scheme imposes small subpacketization and beamforming overheads, is robust under dynamic network conditions, and incurs small finite-SNR performance loss compared with centralized schemes.

\end{abstract}

\begin{IEEEkeywords}
coded caching, MIMO communications, low-subpacketization, dynamic networks
\end{IEEEkeywords}

\section{Introduction}
Wireless networks are under continuous pressure to support increasing volumes of multimedia content~\cite{cisco2018cisco} and pave the way for the emergence of new applications such as wireless immersive viewing~\cite{mahmoodi2021non}. For the efficient delivery of such multimedia content, Maddah-Ali and Niesen proposed the idea of \emph{coded caching} for increasing the data rates by exploiting cache content across the network~\cite{maddah2014fundamental}. In a single-stream downlink network,
coded caching enables boosting the achievable rate by a multiplicative factor proportional to the cumulative cache size in the entire network. This speedup is achieved through
multicasting of carefully created codewords to different groups of users, such that each user can use its cache content to remove unwanted parts 
from the received signal.
Motivated by the growing importance of multi-antenna communications~\cite{rajatheva2020white}, cache-aided \ac{MISO} setting was later explored in~\cite{shariatpanahi2016multi,shariatpanahi2018physical}, where it was revealed that the same coded caching gain could be achieved together with the spatial multiplexing gain. 
This cumulative gain was then shown in~\cite{lampiris2018resolving} to be optimal with the underlying assumptions of uncoded cache placement and one-shot linear data delivery.

Following the introduction of \ac{MISO} coded caching, many later works in the literature addressed its important scaling and performance issues. Notably, in~\cite{tolli2017multi}, the authors showed that using optimized beamformers instead of zero-forcing, one could achieve a better rate for communications at the finite-SNR regime. 
Similarly, the exponentially growing subpacketization issue (i.e., the number of smaller parts each file should be split into) was addressed thoroughly in the literature.
Interestingly, while reducing subpacketization is challenging in single-antenna setups~\cite{yan2018placement},  
%
\ac{MISO} setups allow reducing the subpacketization substantially without affecting the maximum achievable \ac{DoF}~\cite{lampiris2018adding}. 
Of course, this reduced subpacketization modestly decreases the achievable rate, especially at the finite-SNR regime~\cite{salehi2019subpacketization}. Nevertheless, it does not harm the achievable \ac{DoF} and yet enables a much simpler optimized beamformer design~\cite{salehi2020lowcomplexity}.


There exists another critical obstacle preventing the practical implementation of coded caching schemes, especially in dynamic network setups. All the schemes mentioned so far are centralized, i.e., the cache contents of the users are dictated by a central server. However, the server needs prior knowledge of the number of users to indicate what data should be cached at each user during the placement phase. This makes it impossible to implement caching techniques in dynamic networks where the users are allowed to join/leave the network at any moment. One possible solution to this issue is to use fully decentralized schemes, such as~\cite{maddah2015decentralized}, where the cache contents of each user are assumed to be fully random and codewords are built to achieve the caching gain to the maximum possible extent. Unfortunately, the caching gain of such decentralized solutions is comparable with centralized schemes only asymptotically, i.e., when the number of users (or the size of each file) is very large. Hence, the problem remains largely unresolved for practical networks with a moderate number of users.


In this paper, we take a different approach to this problem by introducing a hybrid centralized/decentralized coded caching scheme. In this scheme, the cache contents of users are dictated by a central server, but the users can join/leave the network at any time. Each user is assigned with a \emph{caching profile} that indicates which contents should be cached at its memory. Multiple users may be assigned with the same profile, making the resulting setup similar to shared-cache models where multiple users have access to the same cache storage (cf.~\cite{parrinello2019fundamental}). However, due to the dynamic nature of the network, the \emph{length} of each profile, defined as the number of users associated with that profile, can vary during the time and be independent from the length of other profiles. Then, we introduce a new delivery algorithm, inspired by the low-subpacketization scheme in~\cite{salehi2020lowcomplexity}, for enabling a cumulative caching and spatial multiplexing gain, comparable to a large extent with the ideal case of fully centralized schemes. Of course, this paper is not the first one introducing such a hybrid scheme; a similar approach is introduced in~\cite{jin2019new} for single-antenna setups. However, here, we extended the results to multi-antenna setups and do so with minor subpacketization and beamforming overheads. 

We emphasize that, this paper does \emph{not} aim at proposing an information-theoretic \ac{DoF}-optimal scheme for dynamic networks. Instead, the goal is to provide a practical scheme with an appropriate performance at the finite-SNR regime. In this regard, we leave most of the theoretical analyses of the proposed scheme to the extended version of this paper, and use numerical simulations here to investigate how our new scheme performs in comparison with centralized schemes applied to static network setups. Nevertheless, we provide helpful insights about the underlying reasons for any observation of an improved/deteriorated performance.

In this paper, we use boldface lower- and upper-case letters to denote vectors and matrices, respectively. Sets are shown with calligraphic letters. Similar to vectors, it is assumed that the order of members in a set is important and preserved. $\BA[i,j]$ is the element at row $i$ and column $j$ of matrix $\BA$, and $\Ba[i]$ and $\CA[i]$ are the $i$-th elements of vector $\Ba$ and set $\CA$, respectively.
$[K]$ represents the set of integers $\{1,2,...,K \}$, and $\CA \backslash \CB$ denotes the elements of set $\CA$ which are not in set $\CB$. Other notations are defined as they are used in the text.

\section{System Model}
We consider a \ac{MISO} dynamic setup where a group of single-antenna users request data from a multi-antenna server. The server has spatial \ac{DoF} of $\alpha$.\footnote{The real number of antennas may be larger than the spatial \ac{DoF}. However, we assume the spatial \ac{DoF} at the users and the server is \emph{set} to one and $\alpha$, respectively. In~\cite{tolli2017multi}, it is thoroughly studied how spatial \ac{DoF} value affects the performance of coded caching schemes at finite- and high-SNR regimes.} The users request data from a library $\CF$ of size $N$ files, where the size of each file is $f$ bits. Every user has a cache memory large enough to store a portion $0 < \gamma < 1$ of the entire library. We define the parameter $P$ to be the smallest integer such that $P \gamma$ is also an integer. For example, if $\gamma = \frac{1}{4}$, then, $P = 4$.

The users are allowed to join and leave the network at any time. Upon joining the network, every user $k$ is associated (e.g., randomly) with a profile index $p(k) \in [P]$. The parameter $p(k)$ indicates which subset of data should be cached at the cache memory of user $k$ (more detailed explanation is provided shortly after). After being associated with the profile index $p(k)$, user $k$ starts filling up its cache memory by downloading data from the server. We assume this cache filling process does not affect simultaneously ongoing data delivery processes.\footnote{This is the case, for example, if the cache is filled when excess transmission capacity becomes available at the server (e.g., when the network traffic load is low), or if the users join the network from specific locations with large transmission capacities (e.g., when transmit antennas are located near the entrance door of the application area in an extended reality use case).} After all the required data is downloaded in the cache memory of user $k$, this user is identified by the server to be eligible to participate in upcoming coded caching (CC) data delivery procedures.

Data requests are revealed to the server at specific time intervals. At the beginning of each time interval, a subset of users reveal their requested data from the library $\CF$.\footnote{The reason for assuming that a \emph{subset} of users request data is that in some applications, the requests are made only when the user conditions are changed. For example, consider an extended reality application where the requests are made only when the users move to a different location.} We denote the file requested by user $k$ as $W(k)$. The server then carries out data delivery in two subsequent phases:
\begin{enumerate}
    \item \textbf{CC data delivery}, where the server builds a set of transmission vectors using multi-antenna coded caching techniques and transmits them to eligible users (that have filled their cache memories with the intended data);
    \item \textbf{Unicast data delivery}, where the server transmits all the remaining requested data, not delivered in the CC phase.
\end{enumerate}
The first phase is designed to enable a combined global caching and spatial multiplexing gain, comparable with the ideal case of fully centralized schemes. However, in the unicast phase, only the local caching gain is achieved together with the spatial \ac{DoF}. In this paper, without loss of generality, we consider only a specific request interval and drop the interval index. We assume the whole process is repeated at every time interval after a new subset of users reveal their requests.

The CC delivery phase is based on the RED scheme in~\cite{salehi2020lowcomplexity}. This scheme allows achieving the largest possible combined \ac{DoF} of caching and spatial multiplexing with a very small subpacketization requirement. It also enables implementing high-performance optimized beamformers with very low complexity (using uplink-downlink duality). The same technique is used in this paper to design optimized beamformers for both delivery phases. As discussed earlier, we analyze the finite-SNR performance here, and leave theoretical analysis for the extended version of this paper. In this regards, we use the total required delivery time (over both phases) as our metric. Symmetric rate is then defined as the inverse of the total delivery time.

\textbf{Cache Placement.}
The content cached at each user $k$ is defined by its associated profile $p(k)$. Multiple users may be associated with the same profile, and hence, cache the same content.
In order to indicate which data elements should be stored for each profile, we use a $P \times P$ binary placement matrix $\BV$. The first row of $\BV$ has $\Brt = P \gamma$ consecutive ones (other elements are zero), and the row $1 < p \le P$ is a circular shift of the row $p-1$ to the right by one unit. Given $\BV$, we split each file $W \in \CF$ into $P$ equal-sized \emph{packets} $W_p$. Then, if $\BV[p,p(k)] = 1$ for some $k$ and $p$, user $k$ is instructed to cache packets $W_p$ of every file $W \in \CF$.
\begin{exmp}
\label{exmp:base_network}
Consider a MISO setup where $\gamma = \frac{1}{4}$, and hence, $P=4$ and $\Brt = 1$. Then, the placement matrix is
\begin{equation*}
    \BV = 
     \renewcommand\arraystretch{0.6}\begin{bmatrix}
    1 & 0 & 0 & 0\\
    0 & 1 & 0 & 0\\
    0 & 0 & 1 & 0\\
    0 & 0 & 0 & 1
    \end{bmatrix} \; .
\end{equation*}
So, each file is split into $P=4$ packets, and, for example, if $p(k)=1$ for some user $k$, then it has to cache packet $W_1$ from every file $W \in \CF$.
\end{exmp}


\section{CC Delivery Phase}
After the users reveal their requests, the server first performs a CC delivery phase where it uses coded caching techniques to build a set of codewords and transmit them to eligible users. Assume the number of users associated with profile $p$ and making requests at the considered time instant (i.e., the length of profile $p$) is $\eta_p$. The server then chooses a common parameter $\hat{\eta}$ to indicate the number of users participating in the CC delivery phase. The system level impact of choosing $\hat{\eta}$ is clarified later in this section. Then, for every profile $p$:
\begin{enumerate}
    \item if $\eta_p > \hat{\eta}$, then $\hat{\eta} - \eta_p$ users associated with profile $p$ are selected randomly and excluded from the CC phase;
    \item if $\eta_p \le \hat{\eta}$, then $\hat{\eta} - \eta_p$ \textit{phantom} users are assigned with the profile $p$.
\end{enumerate}
Phantom users are non-existing users that appear when codewords are built and are excluded from the actual delivery~\cite{salehi2020lowcomplexity}. Although phantom users incur some \ac{DoF} loss, they enable a larger beamforming gain that can actually improve the achievable rate, compared with schemes with a larger DoF, at the finite-SNR regime (cf.~\cite{salehi2020lowcomplexity,tolli2017multi}). 

After the above-mentioned process, all profiles will have the same length of $\hat{\eta}$. Then, we consider a \emph{virtual} network with $P$ users, coded caching gain $\Brt = P\gamma$, and spatial DoF of $\Bar{\alpha} = \lceil \frac{\alpha}{\hat{\eta}} \rceil$, where the virtual user $\Brk \in [P]$ has the cache content corresponding to the profile $p = \Brk$ in the original network. For this virtual network, we build the transmission vectors using the RED scheme in~\cite{salehi2020lowcomplexity},\footnote{The RED scheme requires the spatial DoF to be larger than or equal to the caching gain. So, here we assume $\Bar{\alpha} \ge \Brt$. Relaxing this condition is left for the extended version of the paper.} resulting in a total number of $P(P-\Brt)$ transmission vectors $\Bar{\Bx}_{j}^{r}$, where $j \in [P-\Brt]$ and $r \in [P]$. In fact, the virtual network would require $P$ transmission \emph{rounds}, where at each round, $P-\Brt$ individual transmissions are done. The transmission vector $\Bar{\Bx}_{j}^{r}$ is explicitly built as 
\begin{equation}
\label{eq:trans_general_virt_net}
    \Bar{\Bx}_{j}^{r} = \sum_{\Brn \in [\Brt+\Bar{\alpha}]} \BrW_{\Bar{\Bp}_{j}^{r}[\Brn]}^{\Brq}(\Bar{\Bk}_{j}^{r}[\Brn]) \Bar{\Bw}_{\Bar{\CR}_{j}^{r}(\Brn)} \; ,
\end{equation}
where $\Bar{\Bp}_j^r$ and $\Bar{\Bk}_j^r$ are the packet and user index vectors, and $\Bar{\CR}_j^r(\Brn)$ is the interference indicator set used for the $\Brn$-th data term, in the $j$-th transmission of round $r$. Also, $\Bar{\Bw}_{\Bar{\CR}_{j}^{r}(\Brn)}$ is the optimized beamforming vector suppressing data at every user in $\Bar{\CR}_j^r(\Brn)$, $\BrW(\Brk)$ denotes the file requested by the virtual user $\Brk$, and $\Brq$ is the subpacket index which is initialized to one and increased every time a packet appears in a transmission vector. All these parameters are exactly defined in~\cite{salehi2020lowcomplexity}. Specifically, $\Bar{\Bp}_j^r$ and $\Bar{\Bk}_j^r$ are built such that the graphical representation of the transmission vectors follows two perpendicular circular shift operations over a grid. The following example clarifies how the transmission vectors are built for the virtual network.

\begin{exmp}
\label{exmp:virt_network}
Consider the network in Example~\ref{exmp:base_network}, and assume the spatial DoF is $\alpha = 4$. Also, assume $\eta_1 = \eta_2 = 2$ and $\eta_3 = \eta_4 =3$. If we choose either $\hat{\eta} = 2$ or $\hat{\eta} = 3$, the virtual network will have $P = 4$ users, coded caching gain $\Brt = 1$ and spatial DoF $\Bar{\alpha} = 2$. Then, the data delivery will require four rounds, with three transmissions at each round. According to~\cite{salehi2020lowcomplexity}, user and packet index vectors for the first round are
\begin{equation*}
    \begin{aligned}
        &\Bar{\Bk}_1^1 = [\Bar{1}, \Bar{2}, \Bar{3}], \quad \Bar{\Bk}_2^1 = [\Bar{1}, \Bar{3}, \Bar{4}], \quad \Bar{\Bk}_3^1 = [\Bar{1}, \Bar{4}, \Bar{2}] \; , \\
        &\Bar{\Bp}_1^1 = [\Bar{2}, \Bar{1}, \Bar{1}], \quad \Bar{\Bp}_2^1 = [\Bar{3}, \Bar{1}, \Bar{1}], \quad \Bar{\Bp}_3^1 = [\Bar{4}, \Bar{1}, \Bar{1}] \; ,
    \end{aligned}
\end{equation*}
resulting in the transmission vectors
\begin{equation*}
    \label{eq:virt_trans_vectors}
    \begin{aligned}
        \Bar{\Bx}_1^1 &= \BrW_{\Bar{2}}^{\Bar{1}}(\Bar{1}) \Bar{\Bw}_{\Bar{3}} + \BrW_{\Bar{1}}^{\Bar{1}}(\Bar{2}) \Bar{\Bw}_{\Bar{3}} + \BrW_{\Bar{1}}^{\Bar{1}}(\Bar{3}) \Bar{\Bw}_{\Bar{2}} \; , \\
        \Bar{\Bx}_2^1 &= \BrW_{\Bar{3}}^{\Bar{1}}(\Bar{1}) \Bar{\Bw}_{\Bar{4}} + \BrW_{\Bar{1}}^{\Bar{2}}(\Bar{3}) \Bar{\Bw}_{\Bar{4}} + \BrW_{\Bar{1}}^{\Bar{1}}(\Bar{4}) \Bar{\Bw}_{\Bar{3}} \; , \\
        \Bar{\Bx}_3^1 &= \BrW_{\Bar{4}}^{\Bar{1}}(\Bar{1}) \Bar{\Bw}_{\Bar{2}} + \BrW_{\Bar{1}}^{\Bar{2}}(\Bar{4}) \Bar{\Bw}_{\Bar{2}} + \BrW_{\Bar{1}}^{\Bar{2}}(\Bar{2}) \Bar{\Bw}_{\Bar{4}} \; , \\
    \end{aligned}
\end{equation*}
where brackets for interference indicator sets are dropped for notational simplicity. For clarification, we use a tabular view, borrowed from~\cite{salehi2020lowcomplexity}, to graphically represent the transmission vectors in Figure~\ref{fig:subhatnoxor}. In this representation, columns and rows denote user and packet indices, respectively. A darkly shaded cell means the packet index is cached at the user, while a lightly shaded cell indicates that (part of) the packet index is transmitted to the user. As can be seen in Figure~\ref{fig:subhatnoxor}, the transmission vectors in a single round are built using circular shift operations of the lightly shaded cells over non-colored cells of the table, in two perpendicular directions.

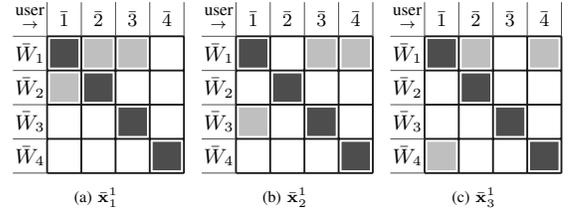
\begin{figure}[t]
    \centering
    \resizebox{0.85\columnwidth}{!}{%
    \begin{subfigure}{.31\columnwidth}
        \centering
        \begin{tikzpicture}[scale = 0.52]
            \begin{scope}<+->;
            \draw[step=1cm,thick,black!90] (1,0) grid (5,4);
            \draw[thin,black!70](0,0)to(1,0);
            \draw[thin,black!70](0,1)to(1,1);
            \draw[thin,black!70](0,2)to(1,2);
            \draw[thin,black!70](0,3)to(1,3);
            \draw[thin,black!70](0,4)to(1,4);
            \draw[thin,black!70](1,4)to(1,5);
            \draw[thin,black!70](2,4)to(2,5);
            \draw[thin,black!70](3,4)to(3,5);
            \draw[thin,black!70](4,4)to(4,5);
            \draw[thin,black!70](5,4)to(5,5);
            \end{scope}
            \begin{scope}
            \FillBlack{5}{2}{2};
            \FillBlack{5}{3}{3};
            \FillBlack{5}{4}{4};
            \FillBlack{5}{5}{5};
            
            \FillGray{5}{2}{3};
            \FillGray{5}{2}{4};
            \FillGray{5}{3}{2};
            
            \PutText{5}{2}{1}{$\BrW_1$};
            \PutText{5}{3}{1}{$\BrW_2$};
            \PutText{5}{4}{1}{$\BrW_3$};
            \PutText{5}{5}{1}{$\BrW_4$};
            \PutText{5}{1}{1}{$\underset{\rightarrow}{\textrm{\smaller{user}}}$};
            \PutText{5}{1}{2}{\smaller{$\Bar{1}$}};
            \PutText{5}{1}{3}{\smaller{$\Bar{2}$}};
            \PutText{5}{1}{4}{\smaller{$\Bar{3}$}};
            \PutText{5}{1}{5}{\smaller{$\Bar{4}$}};
            \end{scope}
        \end{tikzpicture}
        \caption{$\Bar{\Bx}_1^1$}
    \end{subfigure}
    \begin{subfigure}{.31\columnwidth}
        \centering
        \begin{tikzpicture}[scale = 0.52]
            \begin{scope}<+->;
            \draw[step=1cm,thick,black!90] (1,0) grid (5,4);
            \draw[thin,black!70](0,0)to(1,0);
            \draw[thin,black!70](0,1)to(1,1);
            \draw[thin,black!70](0,2)to(1,2);
            \draw[thin,black!70](0,3)to(1,3);
            \draw[thin,black!70](0,4)to(1,4);
            \draw[thin,black!70](1,4)to(1,5);
            \draw[thin,black!70](2,4)to(2,5);
            \draw[thin,black!70](3,4)to(3,5);
            \draw[thin,black!70](4,4)to(4,5);
            \draw[thin,black!70](5,4)to(5,5);
            \end{scope}
            \begin{scope}
            \FillBlack{5}{2}{2};
            \FillBlack{5}{3}{3};
            \FillBlack{5}{4}{4};
            \FillBlack{5}{5}{5};
            
            \FillGray{5}{2}{4};
            \FillGray{5}{2}{5};
            \FillGray{5}{4}{2};
            
            \PutText{5}{2}{1}{$\BrW_1$};
            \PutText{5}{3}{1}{$\BrW_2$};
            \PutText{5}{4}{1}{$\BrW_3$};
            \PutText{5}{5}{1}{$\BrW_4$};
            \PutText{5}{1}{1}{$\underset{\rightarrow}{\textrm{\smaller{user}}}$};
            \PutText{5}{1}{2}{\smaller{$\Bar{1}$}};
            \PutText{5}{1}{3}{\smaller{$\Bar{2}$}};
            \PutText{5}{1}{4}{\smaller{$\Bar{3}$}};
            \PutText{5}{1}{5}{\smaller{$\Bar{4}$}};
            \end{scope}
        \end{tikzpicture}
        \caption{$\Bar{\Bx}_2^1$}
    \end{subfigure}
    \begin{subfigure}{.31\columnwidth}
        \centering
        \begin{tikzpicture}[scale = 0.52]
            \begin{scope}<+->;
            \draw[step=1cm,thick,black!90] (1,0) grid (5,4);
            \draw[thin,black!70](0,0)to(1,0);
            \draw[thin,black!70](0,1)to(1,1);
            \draw[thin,black!70](0,2)to(1,2);
            \draw[thin,black!70](0,3)to(1,3);
            \draw[thin,black!70](0,4)to(1,4);
            \draw[thin,black!70](1,4)to(1,5);
            \draw[thin,black!70](2,4)to(2,5);
            \draw[thin,black!70](3,4)to(3,5);
            \draw[thin,black!70](4,4)to(4,5);
            \draw[thin,black!70](5,4)to(5,5);
            \end{scope}
            \begin{scope}
            \FillBlack{5}{2}{2};
            \FillBlack{5}{3}{3};
            \FillBlack{5}{4}{4};
            \FillBlack{5}{5}{5};
            
            \FillGray{5}{2}{5};
            \FillGray{5}{2}{3};
            \FillGray{5}{5}{2};
            
            \PutText{5}{2}{1}{$\BrW_1$};
            \PutText{5}{3}{1}{$\BrW_2$};
            \PutText{5}{4}{1}{$\BrW_3$};
            \PutText{5}{5}{1}{$\BrW_4$};
            \PutText{5}{1}{1}{$\underset{\rightarrow}{\textrm{\smaller{user}}}$};
            \PutText{5}{1}{2}{\smaller{$\Bar{1}$}};
            \PutText{5}{1}{3}{\smaller{$\Bar{2}$}};
            \PutText{5}{1}{4}{\smaller{$\Bar{3}$}};
            \PutText{5}{1}{5}{\smaller{$\Bar{4}$}};
            \end{scope}
        \end{tikzpicture}
        \caption{$\Bar{\Bx}_3^1$}
    \end{subfigure}
    }
    
    \caption{Graphical representation of transmission vectors for the virtual network}
    \label{fig:subhatnoxor}
\end{figure}
\end{exmp}

After the transmission vectors are built for the virtual network, we need to \emph{elevate} them to be applicable to the original network. We first need to define a few new parameters; $b$ is the remainder of the division of $\alpha$ by $\hat{\eta}$, i.e., $b = \alpha - \hat{\eta} \lfloor \frac{\alpha}{\hat{\eta}} \rfloor$, and the set $\CD_{\Brk}$ consists of the set of user indices in the real network that are associated with the profile $p = \Brk$. Then, we have two elevation mechanisms, depending on the value of $b$.

\noindent \textbf{1)} If $b = 0$, i.e., if $\frac{\alpha}{\hat{\eta}}$ is an integer, every vector $\Bar{\Bx}_j^r$ is elevated into one transmission vector $\Bx_j^r$ for the original network. To do so, each term inside the summation in~\eqref{eq:trans_general_virt_net} is replaced by
\begin{equation*}
\label{eq:elevation_b0}
    \BrW_{\Bar{\Bp}_{j}^{r}[\Brn]}^{\Brq}(\Bar{\Bk}_{j}^{r}[\Brn]) \Bar{\Bw}_{\Bar{\CR}_{j}^{r}(\Brn)} \rightarrow
    \sum_{m \in [\hat{\eta}]} W_{\Bar{\Bp}_{j}^{r}[\Brn]}^{q}(\CD_{\Bar{\Bk}_{j}^{r}[\Brn]}[m]) \Bw_{\CR_j^r(\Brn,m)} \; ,
\end{equation*}
where
\begin{equation}
\label{eq:int_set_elevation}
    \CR_j^r(\Brn,m) = \bigcup_{\Brk \in \Bar{\CR}_j^r(\Brn)} \!\! \CD_{\Brk} \;\; \cup \;\; \CD_{\Bar{\Bk}_{j}^{r}[\Brn]} \;\; \backslash \;\; \{ \CD_{\Bar{\Bk}_{j}^{r}[\Brn]}[m] \} \; .
\end{equation}
In other words, every data term intended for some user $\Brk$ in the virtual network is replaced by $\hat{\eta}$ data terms intended for all users associated with profile $p = \Brk$ in the real network, and the inter-stream interference between these terms is suppressed by beamforming vectors. Note that each interference indicator set $\Bar{\CR}_{j}^{r}(\Brn)$ in the virtual network consists of $\Bar{\alpha} - 1$ virtual users, and hence, using~\eqref{eq:int_set_elevation}, every interference indicator set in the real network will have
    $(\Bar{\alpha}-1)\hat{\eta} + (\hat{\eta}-1) = \Bar{\alpha}\hat{\eta} - 1 = \alpha - 1$
users. So, suppressing the interference with the beamformers is possible as the spatial DoF for the original network is $\alpha$. Summarizing the discussions, if $b=0$, the transmission vectors for the original network are built as
\begin{equation}
    \Bx_j^r = \sum_{\Brn \in [\Brt+\Bar{\alpha}]} \sum_{m \in [\hat{\eta}]} W_{\Bar{\Bp}_{j}^{r}[\Brn]}^{q}(\CD_{\Bar{\Bk}_{j}^{r}[\Brn]}[m]) \Bw_{\CR_j^r(\Brn,m)} \; .
\end{equation}

\noindent \textbf{2)} If $b \neq 0$, i.e., when $\alpha$ is \emph{not} divisible by $\hat{\eta}$, every transmission vector $\Bar{\Bx}_j^r$ is elevated into $\hat{\eta}$ transmission vectors $\Bx_{j,s}^r$, $s \in [\hat{\eta}]$, for the original network. 
For elevation, we first define the \emph{base} and \emph{extended} interference indicator sets as
\begin{equation*}
    \CB_{j,s}^{r,b}(\Brn,m) = \bigcup_{\Brk \in \Bar{\CR}_j^r[\Brn]} \!\! \CD_{\Brk} \;\; \bigcup_{i \in \CE_s^b(\hat{\eta})} \{ \CD_{\Bar{\Bk}_{j}^{r}[\Brt + \Bar{\alpha}]}[i] \}
\end{equation*}
and
\begin{equation*}
    \CR_{j,s}^{r,b}(\Brn,m) \! = \!
    \begin{cases}
    & \!\!\!\!\!\! \CB_{j,s}^{r,b}(\Brn,m) \cup \CD_{\Bar{\Bk}_{j}^{r}[\Brn]} \backslash \{ \CD_{\Bar{\Bk}_{j}^{r}[\Brn]}[m] \} \quad \! \Brn < \Brt+\Bar{\alpha}\\
    & \!\!\!\!\!\! \CB_{j,s}^{r,b}(\Brn,m) \backslash \{\CD_{\Bar{\Bk}_{j}^{r}[\Brn]}[m]\} \quad \quad \quad \quad \;\; \Brn = \Brt+\Bar{\alpha}
    \end{cases}
\end{equation*}
respectively, where $\CE_s^b(\hat{\eta})$ denotes the first $b$ elements of $[\hat{\eta}]$, after they are circularly shifted to the left by $s$ units (recall that the order of members in a set is preserved). In other words,
\begin{equation}
    \CE_s^b(\hat{\eta}) = \big( ([b] + s - 2) \mod \hat{\eta} \big) + 1 \; .
\end{equation}
Then, the elevated transmission vectors are built as
\begin{equation}
    \Bx_{j,s}^r = \sum_{\Brn \in [\Brt+\Bar{\alpha}]} \sum_{m \in [\hat{\eta}]} W_{\Bar{\Bp}_{j}^{r}[\Brn]}^{q}(\CD_{\Bar{\Bk}_{j}^{r}[\Brn]}[m]) \Bw_{\CR_{j,s}^{r,b}(\Brn,m)} \; .
\end{equation}
%
As an explanation, from the $\Brt + \Bar{\alpha}$ total data terms within the transmission vector $\Bar{\Bx}_j^r$, we substitute each of the first $\Brt + \Bar{\alpha}-1$ terms with $\hat{\eta}$ terms for the real network (as we did for the case $b=0$). However, the last data term is only replaced by $b$ terms, out of the $\hat{\eta}$ options, chosen through a circular shift operation (implemented using $\CE_s^b(\hat{\eta})$ in the equations). The following example clarifies the elevation process for both cases of $b = 0$ and $b \neq 0$, and enlightens the effect of phantom users. 

\begin{exmp}
\label{exmp:base_net_cc_perfect}
Consider the same network in Example~\ref{exmp:base_network}, with the profile lengths mentioned in Example~\ref{exmp:virt_network}. Let us assume $\CD_{\Bar{1}} = [1,2]$, $\CD_{\Bar{2}} = [3,4]$, $\CD_{\Bar{3}} = [5,6,7]$, and $\CD_{\Bar{4}} = [8,9,10]$. For this network, we find transmission vectors resulting from the elevation of the vectors in~\eqref{eq:virt_trans_vectors}, for both cases of $\hat{\eta} = 2,3$.

If $\hat{\eta} = 2$, no phantom users are needed. However, we should exclude two users, associated with profiles $p=3,4$, from the CC delivery phase. Assume users 7 and 10 are excluded. As $b=0$, transmission vectors $\Bar{\Bx}_1^1$ and $\Bar{\Bx}_1^2$ in~\eqref{eq:virt_trans_vectors} are elevated as
\begin{equation*}
    \begin{aligned}
        \Bx_1^1 =& W_2^1(1) \Bw_{5,6,2} + W_2^1(2) \Bw_{5,6,1} + W_1^1(3) \Bw_{5,6,4} \\
        &+ W_1^1(4) \Bw_{5,6,3} + W_1^1(5) \Bw_{3,4,6} + W_1^1(6) \Bw_{3,4,5} \; , \\
        \Bx_2^1 =& W_3^1(1) \Bw_{8,9,2} + W_3^1(2) \Bw_{8,9,1} + W_1^2(5) \Bw_{8,9,6} \\
        &+ W_1^2(6) \Bw_{8,9,5} + W_1^1(8) \Bw_{5,6,9} + W_1^1(9) \Bw_{5,6,8} \; , \\
    \end{aligned}
\end{equation*}
and their graphical representation will be as shown in Figure~\ref{fig:elevated_b0}. Comparing Figures~\ref{fig:subhatnoxor} and~\ref{fig:elevated_b0}, we can see that the elevation mechanism stretches the transmission vectors horizontally.

\begin{figure}[t]

\centering

\begin{subfigure}{\columnwidth}
    \centering
    \resizebox{0.86\columnwidth}{!}{%
    \begin{subfigure}{.65\columnwidth}
        \centering
        \begin{tikzpicture}[scale = 0.52]
            \begin{scope}<+->;
            \draw[step=1cm,thick,black!90] (1,0) grid (9,4);
            \draw[thin,black!70](0,0)to(1,0);
            \draw[thin,black!70](0,1)to(1,1);
            \draw[thin,black!70](0,2)to(1,2);
            \draw[thin,black!70](0,3)to(1,3);
            \draw[thin,black!70](0,4)to(1,4);
            \draw[thin,black!70](1,4)to(1,5);
            \draw[thin,black!70](2,4)to(2,5);
            \draw[thin,black!70](3,4)to(3,5);
            \draw[thin,black!70](4,4)to(4,5);
            \draw[thin,black!70](5,4)to(5,5);
            \draw[thin,black!70](6,4)to(6,5);
            \draw[thin,black!70](7,4)to(7,5);
            \draw[thin,black!70](8,4)to(8,5);
            \draw[thin,black!70](9,4)to(9,5);
            \end{scope}
            \begin{scope}
            \FillBlack{5}{2}{2};
            \FillBlack{5}{2}{3};
            \FillBlack{5}{3}{4};
            \FillBlack{5}{3}{5};
            \FillBlack{5}{4}{6};
            \FillBlack{5}{4}{7};
            \FillBlack{5}{5}{8};
            \FillBlack{5}{5}{9};
            
            \FillGray{5}{2}{4};
            \FillGray{5}{2}{5};
            \FillGray{5}{2}{6};
            \FillGray{5}{2}{7};
            \FillGray{5}{3}{2};
            \FillGray{5}{3}{3};
            
            \PutText{5}{2}{1}{$W_1$};
            \PutText{5}{3}{1}{$W_2$};
            \PutText{5}{4}{1}{$W_3$};
            \PutText{5}{5}{1}{$W_4$};
            \PutText{5}{1}{1}{$\underset{\rightarrow}{\textrm{\smaller{user}}}$};
            \PutText{5}{1}{2}{\smaller{${1}$}};
            \PutText{5}{1}{3}{\smaller{${2}$}};
            \PutText{5}{1}{4}{\smaller{${3}$}};
            \PutText{5}{1}{5}{\smaller{${4}$}};
            \PutText{5}{1}{6}{\smaller{${5}$}};
            \PutText{5}{1}{7}{\smaller{${6}$}};
            \PutText{5}{1}{8}{\smaller{${8}$}};
            \PutText{5}{1}{9}{\smaller{${9}$}};
            \end{scope}
        \end{tikzpicture}
        \caption*{${\Bx}_1^1$ - CC Delivery}
    \end{subfigure}
    \begin{subfigure}{.31\columnwidth}
        \centering
        \begin{tikzpicture}[scale = 0.52]
            \begin{scope}<+->;
            \draw[step=1cm,thick,black!90] (1,0) grid (3,4);
            \draw[thin,black!70](0,0)to(1,0);
            \draw[thin,black!70](0,1)to(1,1);
            \draw[thin,black!70](0,2)to(1,2);
            \draw[thin,black!70](0,3)to(1,3);
            \draw[thin,black!70](0,4)to(1,4);
            \draw[thin,black!70](1,4)to(1,5);
            \draw[thin,black!70](2,4)to(2,5);
            \draw[thin,black!70](3,4)to(3,5);
            \end{scope}
            \begin{scope}
            \FillBlack{5}{4}{2};
            \FillBlack{5}{5}{3};

            \PutText{5}{2}{1}{$W_1$};
            \PutText{5}{3}{1}{$W_2$};
            \PutText{5}{4}{1}{$W_3$};
            \PutText{5}{5}{1}{$W_4$};
            \PutText{5}{1}{1}{$\underset{\rightarrow}{\textrm{\smaller{user}}}$};
            \PutText{5}{1}{2}{\smaller{$7$}};
            \PutText{5}{1}{3}{\smaller{$10$}};
            \end{scope}
        \end{tikzpicture}
        \caption*{Left for Unicast}
    \end{subfigure}
    }
\end{subfigure}

\begin{subfigure}{\columnwidth}
    \centering
    \resizebox{0.86\columnwidth}{!}{%
    \begin{subfigure}{.65\columnwidth}
        \centering
        \begin{tikzpicture}[scale = 0.52]
            \begin{scope}<+->;
            \draw[step=1cm,thick,black!90] (1,0) grid (9,4);
            \draw[thin,black!70](0,0)to(1,0);
            \draw[thin,black!70](0,1)to(1,1);
            \draw[thin,black!70](0,2)to(1,2);
            \draw[thin,black!70](0,3)to(1,3);
            \draw[thin,black!70](0,4)to(1,4);
            \draw[thin,black!70](1,4)to(1,5);
            \draw[thin,black!70](2,4)to(2,5);
            \draw[thin,black!70](3,4)to(3,5);
            \draw[thin,black!70](4,4)to(4,5);
            \draw[thin,black!70](5,4)to(5,5);
            \draw[thin,black!70](6,4)to(6,5);
            \draw[thin,black!70](7,4)to(7,5);
            \draw[thin,black!70](8,4)to(8,5);
            \draw[thin,black!70](9,4)to(9,5);
            \end{scope}
            \begin{scope}
            \FillBlack{5}{2}{2};
            \FillBlack{5}{2}{3};
            \FillBlack{5}{3}{4};
            \FillBlack{5}{3}{5};
            \FillBlack{5}{4}{6};
            \FillBlack{5}{4}{7};
            \FillBlack{5}{5}{8};
            \FillBlack{5}{5}{9};
            
            \FillGray{5}{2}{6};
            \FillGray{5}{2}{7};
            \FillGray{5}{2}{8};
            \FillGray{5}{2}{9};
            \FillGray{5}{4}{2};
            \FillGray{5}{4}{3};
            
            \PutText{5}{2}{1}{$W_1$};
            \PutText{5}{3}{1}{$W_2$};
            \PutText{5}{4}{1}{$W_3$};
            \PutText{5}{5}{1}{$W_4$};
            \PutText{5}{1}{1}{$\underset{\rightarrow}{\textrm{\smaller{user}}}$};
            \PutText{5}{1}{2}{\smaller{${1}$}};
            \PutText{5}{1}{3}{\smaller{${2}$}};
            \PutText{5}{1}{4}{\smaller{${3}$}};
            \PutText{5}{1}{5}{\smaller{${4}$}};
            \PutText{5}{1}{6}{\smaller{${5}$}};
            \PutText{5}{1}{7}{\smaller{${6}$}};
            \PutText{5}{1}{8}{\smaller{${8}$}};
            \PutText{5}{1}{9}{\smaller{${9}$}};
            \end{scope}
        \end{tikzpicture}
        \caption*{${\Bx}_2^1$ - CC Delivery}
    \end{subfigure}
    \begin{subfigure}{.31\columnwidth}
        \centering
        \begin{tikzpicture}[scale = 0.48]
            \begin{scope}<+->;
            \draw[step=1cm,thick,black!90] (1,0) grid (3,4);
            \draw[thin,black!70](0,0)to(1,0);
            \draw[thin,black!70](0,1)to(1,1);
            \draw[thin,black!70](0,2)to(1,2);
            \draw[thin,black!70](0,3)to(1,3);
            \draw[thin,black!70](0,4)to(1,4);
            \draw[thin,black!70](1,4)to(1,5);
            \draw[thin,black!70](2,4)to(2,5);
            \draw[thin,black!70](3,4)to(3,5);
            \end{scope}
            \begin{scope}
            \FillBlack{5}{4}{2};
            \FillBlack{5}{5}{3};

            \PutText{5}{2}{1}{$W_1$};
            \PutText{5}{3}{1}{$W_2$};
            \PutText{5}{4}{1}{$W_3$};
            \PutText{5}{5}{1}{$W_4$};
            \PutText{5}{1}{1}{$\underset{\rightarrow}{\textrm{\smaller{user}}}$};
            \PutText{5}{1}{2}{\smaller{$7$}};
            \PutText{5}{1}{3}{\smaller{$10$}};
            \end{scope}
        \end{tikzpicture}
        \caption*{Left for Unicast}
    \end{subfigure}
    }
\end{subfigure}

\caption{Elevated versions of $\Bar{\Bx}_1^1$ (top) and $\Bar{\Bx}_2^1$ (bottom) - The case $b=0$}
\label{fig:elevated_b0}

\end{figure}
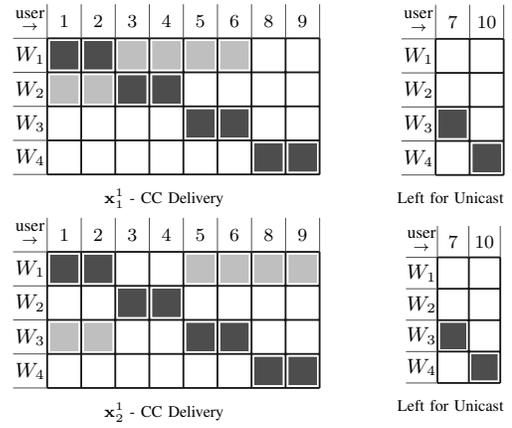

Now, let us assume $\hat{\eta} = 3$. In this case, no user is excluded from the CC delivery phase, but two phantom users $\Check{1}$ and $\Check{2}$ are associated with profiles $p=1$ and $p=2$, respectively. As $b=1$, each transmission vector for the virtual network is elevated into $\hat{\eta} = 3$ vectors for the real network. Here we only mention the vectors resulting from $\Bar{\Bx}_1^1$, which are built as
\begin{equation}
\label{eq:trans_vect_b_neq_0_phantom}
    \begin{aligned}
        &\Bx_{1,1}^1 \! = \! W_2^1(1) \Bw_{2,\Check{1},5} \! + \! W_2^1(2) \Bw_{1,\Check{1},5} \! + \! W_2^1(\Check{1}) \Bw_{1,2,5} + \\
        & \; \; \; W_1^1(3) \Bw_{4,\Check{2},5} \! + \! W_1^1(4) \Bw_{3,\Check{2},5} \! + \! W_1^1(\Check{2}) \Bw_{3,4,5} \! + \! W_1^1(5) \Bw_{3,4,\Check{2}} , \\
        &\Bx_{1,2}^1 \! = \! W_2^2(1) \Bw_{2,\Check{1},6} \! + \! W_2^2(2) \Bw_{1,\Check{1},6} \! + \! W_2^2(\Check{1}) \Bw_{1,2,6} + \\
        & \; \; \; W_1^2(3) \Bw_{4,\Check{2},6} \! + \! W_1^2(4) \Bw_{3,\Check{2},6} \! + \! W_1^2(\Check{2}) \Bw_{3,4,6} \! + \! W_1^1(6) \Bw_{3,4,\Check{2}} , \\
        &\Bx_{1,3}^1 \! = \! W_2^3(1) \Bw_{2,\Check{1},7} \! + \! W_2^3(2) \Bw_{1,\Check{1},7} \! + \! W_2^3(\Check{1}) \Bw_{1,2,7} + \\
        & \; \; \; W_1^3(3) \Bw_{4,\Check{2},7} \! + \! W_1^3(4) \Bw_{3,\Check{2},7} \! + \! W_1^3(\Check{2}) \Bw_{3,4,7} \! + \! W_1^1(7) \Bw_{3,4,\Check{2}} . \\
    \end{aligned}
\end{equation}
The graphical representation of this elevation mechanism is shown in Figure~\ref{fig:elevated_b1}, where the columns representing phantom users are hatched for better clarification. Comparing with the base transmission vectors for the virtual network in Figure~\ref{fig:subhatnoxor}, the elevation mechanism works through horizontal stretching followed by an extra circular shift operation over a block of user indices (users 5,6,7 in this example). Finally, we should note that the effect of phantom users should be removed before the real transmission. To do so, we simply remove phantom indices from the transmission vectors. So, for example, instead of $\Bx_{1,1}^1$ in~\eqref{eq:trans_vect_b_neq_0_phantom}, we transmit
\begin{equation*}
\begin{aligned}
    \Bx_{1,1}^1 =& W_2^1(1) \Bw_{2,5} + W_2^1(2) \Bw_{1,5} + W_1^1(3) \Bw_{4,5} \\
    & + W_1^1(4) \Bw_{3,5} + W_1^1(5) \Bw_{3,4} \; . \\
\end{aligned}
\end{equation*}

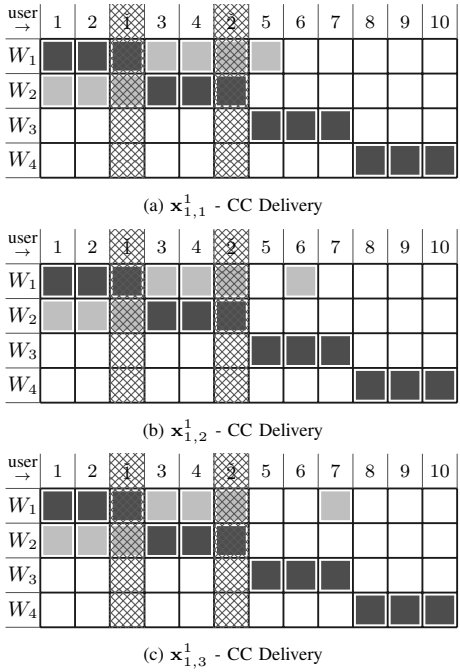
\begin{figure}[tp]
    \centering
    \begin{subfigure}{\columnwidth}
        \centering
        \resizebox{0.7\columnwidth}{!}{%
        \begin{tikzpicture}[scale = 0.52]
            \begin{scope}<+->;
            \draw[step=1cm,thick,black!90] (1,0) grid (13,4);
            \draw[thin,black!70](0,0)to(1,0);
            \draw[thin,black!70](0,1)to(1,1);
            \draw[thin,black!70](0,2)to(1,2);
            \draw[thin,black!70](0,3)to(1,3);
            \draw[thin,black!70](0,4)to(1,4);
            \draw[thin,black!70](1,4)to(1,5);
            \draw[thin,black!70](2,4)to(2,5);
            \draw[thin,black!70](3,4)to(3,5);
            \draw[thin,black!70](4,4)to(4,5);
            \draw[thin,black!70](5,4)to(5,5);
            \draw[thin,black!70](6,4)to(6,5);
            \draw[thin,black!70](7,4)to(7,5);
            \draw[thin,black!70](8,4)to(8,5);
            \draw[thin,black!70](9,4)to(9,5);
            \draw[thin,black!70](10,4)to(10,5);
            \draw[thin,black!70](11,4)to(11,5);
            \draw[thin,black!70](12,4)to(12,5);
            \draw[thin,black!70](13,4)to(13,5);
            \end{scope}
            \begin{scope}
            \FillBlack{5}{2}{2};
            \FillBlack{5}{2}{3};
            \FillBlack{5}{2}{4};
            \FillBlack{5}{3}{5};
            \FillBlack{5}{3}{6};
            \FillBlack{5}{3}{7};
            \FillBlack{5}{4}{8};
            \FillBlack{5}{4}{9};
            \FillBlack{5}{4}{10};
            \FillBlack{5}{5}{11};
            \FillBlack{5}{5}{12};
            \FillBlack{5}{5}{13};
            
            \FillGray{5}{2}{5};
            \FillGray{5}{2}{6};
            \FillGray{5}{2}{7};
            \FillGray{5}{2}{8};
            \FillGray{5}{3}{2};
            \FillGray{5}{3}{3};
            \FillGray{5}{3}{4};
            
            \FillHatch{5}{1}{4};
            \FillHatch{5}{2}{4};
            \FillHatch{5}{3}{4};
            \FillHatch{5}{4}{4};
            \FillHatch{5}{5}{4};
            \FillHatch{5}{1}{7};
            \FillHatch{5}{2}{7};
            \FillHatch{5}{3}{7};
            \FillHatch{5}{4}{7};
            \FillHatch{5}{5}{7};
            
            \PutText{5}{2}{1}{$W_1$};
            \PutText{5}{3}{1}{$W_2$};
            \PutText{5}{4}{1}{$W_3$};
            \PutText{5}{5}{1}{$W_4$};
            \PutText{5}{1}{1}{$\underset{\rightarrow}{\textrm{\smaller{user}}}$};
            \PutText{5}{1}{2}{\smaller{${1}$}};
            \PutText{5}{1}{3}{\smaller{${2}$}};
            \PutText{5}{1}{4}{\smaller{$\check{1}$}};
            \PutText{5}{1}{5}{\smaller{${3}$}};
            \PutText{5}{1}{6}{\smaller{${4}$}};
            \PutText{5}{1}{7}{\smaller{$\check{2}$}};
            \PutText{5}{1}{8}{\smaller{${5}$}};
            \PutText{5}{1}{9}{\smaller{${6}$}};
            \PutText{5}{1}{10}{\smaller{${7}$}};
            \PutText{5}{1}{11}{\smaller{${8}$}};
            \PutText{5}{1}{12}{\smaller{${9}$}};
            \PutText{5}{1}{13}{\smaller{${10}$}};
            \end{scope}
        \end{tikzpicture}
        }
        \caption{${\Bx}_{1,1}^1$ - CC Delivery}
    \end{subfigure}
    \begin{subfigure}{\columnwidth}
        \centering
        \resizebox{0.7\columnwidth}{!}{%
        \begin{tikzpicture}[scale = 0.52]
            \begin{scope}<+->;
            \draw[step=1cm,thick,black!90] (1,0) grid (13,4);
            \draw[thin,black!70](0,0)to(1,0);
            \draw[thin,black!70](0,1)to(1,1);
            \draw[thin,black!70](0,2)to(1,2);
            \draw[thin,black!70](0,3)to(1,3);
            \draw[thin,black!70](0,4)to(1,4);
            \draw[thin,black!70](1,4)to(1,5);
            \draw[thin,black!70](2,4)to(2,5);
            \draw[thin,black!70](3,4)to(3,5);
            \draw[thin,black!70](4,4)to(4,5);
            \draw[thin,black!70](5,4)to(5,5);
            \draw[thin,black!70](6,4)to(6,5);
            \draw[thin,black!70](7,4)to(7,5);
            \draw[thin,black!70](8,4)to(8,5);
            \draw[thin,black!70](9,4)to(9,5);
            \draw[thin,black!70](10,4)to(10,5);
            \draw[thin,black!70](11,4)to(11,5);
            \draw[thin,black!70](12,4)to(12,5);
            \draw[thin,black!70](13,4)to(13,5);
            \end{scope}
            \begin{scope}
            \FillBlack{5}{2}{2};
            \FillBlack{5}{2}{3};
            \FillBlack{5}{2}{4};
            \FillBlack{5}{3}{5};
            \FillBlack{5}{3}{6};
            \FillBlack{5}{3}{7};
            \FillBlack{5}{4}{8};
            \FillBlack{5}{4}{9};
            \FillBlack{5}{4}{10};
            \FillBlack{5}{5}{11};
            \FillBlack{5}{5}{12};
            \FillBlack{5}{5}{13};
            
            \FillGray{5}{2}{5};
            \FillGray{5}{2}{6};
            \FillGray{5}{2}{7};
            \FillGray{5}{2}{9};
            \FillGray{5}{3}{2};
            \FillGray{5}{3}{3};
            \FillGray{5}{3}{4};
            
            \FillHatch{5}{1}{4};
            \FillHatch{5}{2}{4};
            \FillHatch{5}{3}{4};
            \FillHatch{5}{4}{4};
            \FillHatch{5}{5}{4};
            \FillHatch{5}{1}{7};
            \FillHatch{5}{2}{7};
            \FillHatch{5}{3}{7};
            \FillHatch{5}{4}{7};
            \FillHatch{5}{5}{7};
            
            \PutText{5}{2}{1}{$W_1$};
            \PutText{5}{3}{1}{$W_2$};
            \PutText{5}{4}{1}{$W_3$};
            \PutText{5}{5}{1}{$W_4$};
            \PutText{5}{1}{1}{$\underset{\rightarrow}{\textrm{\smaller{user}}}$};
            \PutText{5}{1}{2}{\smaller{${1}$}};
            \PutText{5}{1}{3}{\smaller{${2}$}};
            \PutText{5}{1}{4}{\smaller{$\check{1}$}};
            \PutText{5}{1}{5}{\smaller{${3}$}};
            \PutText{5}{1}{6}{\smaller{${4}$}};
            \PutText{5}{1}{7}{\smaller{$\check{2}$}};
            \PutText{5}{1}{8}{\smaller{${5}$}};
            \PutText{5}{1}{9}{\smaller{${6}$}};
            \PutText{5}{1}{10}{\smaller{${7}$}};
            \PutText{5}{1}{11}{\smaller{${8}$}};
            \PutText{5}{1}{12}{\smaller{${9}$}};
            \PutText{5}{1}{13}{\smaller{${10}$}};
            \end{scope}
        \end{tikzpicture}
        }
        \caption{${\Bx}_{1,2}^1$ - CC Delivery}
    \end{subfigure}
    \begin{subfigure}{\columnwidth}
        \centering
        \resizebox{0.7\columnwidth}{!}{%
        \begin{tikzpicture}[scale = 0.52]
            \begin{scope}<+->;
            \draw[step=1cm,thick,black!90] (1,0) grid (13,4);
            \draw[thin,black!70](0,0)to(1,0);
            \draw[thin,black!70](0,1)to(1,1);
            \draw[thin,black!70](0,2)to(1,2);
            \draw[thin,black!70](0,3)to(1,3);
            \draw[thin,black!70](0,4)to(1,4);
            \draw[thin,black!70](1,4)to(1,5);
            \draw[thin,black!70](2,4)to(2,5);
            \draw[thin,black!70](3,4)to(3,5);
            \draw[thin,black!70](4,4)to(4,5);
            \draw[thin,black!70](5,4)to(5,5);
            \draw[thin,black!70](6,4)to(6,5);
            \draw[thin,black!70](7,4)to(7,5);
            \draw[thin,black!70](8,4)to(8,5);
            \draw[thin,black!70](9,4)to(9,5);
            \draw[thin,black!70](10,4)to(10,5);
            \draw[thin,black!70](11,4)to(11,5);
            \draw[thin,black!70](12,4)to(12,5);
            \draw[thin,black!70](13,4)to(13,5);
            \end{scope}
            \begin{scope}
            \FillBlack{5}{2}{2};
            \FillBlack{5}{2}{3};
            \FillBlack{5}{2}{4};
            \FillBlack{5}{3}{5};
            \FillBlack{5}{3}{6};
            \FillBlack{5}{3}{7};
            \FillBlack{5}{4}{8};
            \FillBlack{5}{4}{9};
            \FillBlack{5}{4}{10};
            \FillBlack{5}{5}{11};
            \FillBlack{5}{5}{12};
            \FillBlack{5}{5}{13};
            
            \FillGray{5}{2}{5};
            \FillGray{5}{2}{6};
            \FillGray{5}{2}{7};
            \FillGray{5}{2}{10};
            \FillGray{5}{3}{2};
            \FillGray{5}{3}{3};
            \FillGray{5}{3}{4};
            
            \FillHatch{5}{1}{4};
            \FillHatch{5}{2}{4};
            \FillHatch{5}{3}{4};
            \FillHatch{5}{4}{4};
            \FillHatch{5}{5}{4};
            \FillHatch{5}{1}{7};
            \FillHatch{5}{2}{7};
            \FillHatch{5}{3}{7};
            \FillHatch{5}{4}{7};
            \FillHatch{5}{5}{7};
            
            \PutText{5}{2}{1}{$W_1$};
            \PutText{5}{3}{1}{$W_2$};
            \PutText{5}{4}{1}{$W_3$};
            \PutText{5}{5}{1}{$W_4$};
            \PutText{5}{1}{1}{$\underset{\rightarrow}{\textrm{\smaller{user}}}$};
            \PutText{5}{1}{2}{\smaller{${1}$}};
            \PutText{5}{1}{3}{\smaller{${2}$}};
            \PutText{5}{1}{4}{\smaller{$\check{1}$}};
            \PutText{5}{1}{5}{\smaller{${3}$}};
            \PutText{5}{1}{6}{\smaller{${4}$}};
            \PutText{5}{1}{7}{\smaller{$\check{2}$}};
            \PutText{5}{1}{8}{\smaller{${5}$}};
            \PutText{5}{1}{9}{\smaller{${6}$}};
            \PutText{5}{1}{10}{\smaller{${7}$}};
            \PutText{5}{1}{11}{\smaller{${8}$}};
            \PutText{5}{1}{12}{\smaller{${9}$}};
            \PutText{5}{1}{13}{\smaller{${10}$}};
            \end{scope}
        \end{tikzpicture}
        }
        \caption{${\Bx}_{1,3}^1$ - CC Delivery}
    \end{subfigure}

    \caption{Elevated versions of $\Bar{\Bx}_1^1$ - The case $b \neq 0$}
    \label{fig:elevated_b1}
\end{figure}

\end{exmp}

As can be seen in Example~\ref{exmp:base_net_cc_perfect}, the choice of $\hat{\eta}$ can heavily affect both performance and complexity metrics such as the required subpacketization and number of transmissions. A smaller $\hat{\eta}$ reduces the chance of requiring phantom users, but more real users are excluded from the CC delivery phase, and hence, the achievable coded caching gain is reduced. On the other hand, while with a larger $\hat{\eta}$ we can include more users in the CC delivery phase, in specific transmissions, the achievable \ac{DoF} can become too small after the effect of phantom users is removed (in Example~\ref{exmp:base_net_cc_perfect}, $\Bx_1^1$ has a DoF of six, but for $\Bx_{1,s}^1$, $s\in[3]$ the DoF is reduced to five). In fact, if the distribution of profile lengths $\eta_p$ and the choice of $\hat{\eta}$ are such that we need many phantom users, the achievable DoF in some transmissions may fall below the spatial \ac{DoF} $\alpha$ (i.e., we lose \ac{DoF} by using coded caching techniques). One solution to this issue is to check the achievable \ac{DoF} at every transmission vector after removing the effect of phantom users, and avoid the transmission if the \ac{DoF} falls below $\alpha$. The data terms intended to be sent with such transmission vectors are then transmitted during the subsequent unicast delivery phase.


\section{Unicast Delivery Phase}
During the unicast delivery phase, we fulfill the request of users excluded from the CC delivery phase. Meantime, we also transmit any other data term not sent during the CC delivery phase. This includes, for example, data terms in transmission vectors with a \ac{DoF} value smaller than $\alpha$ (due to the presence of phantom users), or data terms re-transmitted because of channel errors. We use a greedy mechanism (which is not necessarily optimal) to achieve an appropriate spatial \ac{DoF} for every transmission vector in this phase. To do so, we first split the files requested by users excluded from the CC delivery phase into the same number of subpackets we needed in the CC delivery phase (this subpacketization value is discussed shortly after). Then, for every user $k$, we define $u(k)$ to be the number of subpackets that should be delivered to user $k$, and sort users by their $u(k)$ values in the descending order. Now, we select $\max\{\alpha,U\}$ first users, where $U$ is the total number of users for which $u(k) > 0$, and deliver one subpacket to each selected user, with a single transmission vector. The inter-stream interference in the transmission vector is suppressed by beamforming vectors (which is possible as the spatial \ac{DoF} is $\alpha$). Finally, we update $u(k)$ values and repeat the same procedure, until all the remaining data terms are transmitted.

\begin{exmp}
Consider the network in Example~\ref{exmp:base_net_cc_perfect}, and assume $\hat{\eta} = 2$. In this case, we don't need any phantom users, but the users 7 and 10 are excluded from the CC delivery phase (as also shown in Figure~\ref{fig:elevated_b0}). As will be shown later, the total subpacketization in this case is $P(\Brt + \Bar{\alpha}) = 12$. However, in the placement phase, each file is already split into $P=4$ packets, and one packet is cached at each user. So, during the delivery, we need to further split each packet into $\frac{12}{4} = 3$ subpackets and deliver nine subpackets to each of these users. So, $u(k)=9$ for $k=7,10$ and zero otherwise, and $U=2$. Then, we need nine transmission vectors, where each vector delivers one subpacket to each of the users 7 and 10. For example, the first and second vectors are built as $W_1^1(7)\Bw_{10} + W_1^1(10) \Bw_7$ and $W_1^2(7)\Bw_{10} + W_1^2(10) \Bw_7$, respectively.
\end{exmp}


\section{Performance Analysis}
As stated earlier, the goal of this paper is \emph{not} to provide a theoretically optimal scheme. Instead, we aim at proposing a practical coded caching solution for dynamic networks where the users are allowed to join and leave the network. In this regard, our scheme has a small subpacketization requirement and enables a very simple optimized beamformer design using uplink-downlink duality~\cite{salehi2020lowcomplexity}, and hence, can be applied to networks with a large ($O(10^2)$) number of users. Also, as we will verify through simulations, it is able to provide a comparable performance with the non-dynamic scenario.

The subpacketization requirement of our scheme depends on the value of $b$. If $b=0$ (i.e., if $\alpha$ is divisible by $\hat{\eta}$), the total subpacketization is $P(\Brt + \Bar{\alpha})$. This simply follows the fact that in this case, every transmission vector for the virtual network is elevated into exactly one vector for the original network, and no extra splitting is required. However, if $b \neq 0$, the scheme needs a larger subpacketization of $P(\hat{\eta} \Brt+\alpha)$. Analyzing the required subpacketization in this case is lengthy due to the extra circular shift operation in the scheme and removed here due to the lack of space. It can be seen that in both cases, the required subpacketization is pretty small and grows linearly with the total number of users in the CC delivery phase. 

\begin{figure}[t]
    \centering
    \resizebox{0.7\columnwidth}{!}{%
    
    \begin{tikzpicture}

    \begin{axis}
    [
    axis lines = left,
    xlabel = \smaller {SNR [dB]},
    ylabel = \smaller {Symmetric Rate [nats/s]},
    ylabel near ticks,
    legend pos = north west,
    ticklabel style={font=\smaller},
    grid=both,
    major grid style={line width=.2pt,draw=gray!30},
    ]
    
    \addplot[black,thick]
    table[y=A5,x=SNR-dB]{Data/L12.tex};
    \addlegendentry{\smaller Uniform, $\hat{\eta} = 5$}
    
    \addplot[mark = x,gray]
    table[y=B3,x=SNR-dB]{Data/L12.tex};
    \addlegendentry{\smaller Scenario 1, $\hat{\eta} = 3$}
    
    \addplot[mark = triangle,gray]
    table[y=B5,x=SNR-dB]{Data/L12.tex};
    \addlegendentry{\smaller Scenario 1, $\hat{\eta} = 5$}
    
    \addplot[gray]
    table[y=B7,x=SNR-dB]{Data/L12.tex};
    \addlegendentry{\smaller Scenario 1, $\hat{\eta} = 7$}
    
    \addplot[mark = x,black,dashed]
    table[y=D5,x=SNR-dB]{Data/L12.tex};
    \addlegendentry{\smaller Scenario 3, $\hat{\eta} = 5$}
    
    \addplot[mark = triangle,black,dashed]
    table[y=D9,x=SNR-dB]{Data/L12.tex};
    \addlegendentry{\smaller Scenario 3, $\hat{\eta} = 9$}
    
    \addplot[black,dashed]
    table[y=D10,x=SNR-dB]{Data/L12.tex};
    \addlegendentry{\smaller Scenario 3, $\hat{\eta} = 10$}
    
    \addplot[gray,thick]
    table[y=NoCC,x=SNR-dB]{Data/L12.tex};
    \addlegendentry{\smaller No Coded Caching}

    \end{axis}

    \end{tikzpicture}
    }
    \caption{Performance comparison for various $\hat{\eta}$ values, scenarios 1 and 3}
    \label{fig:L12-ScenarioB}
\end{figure}
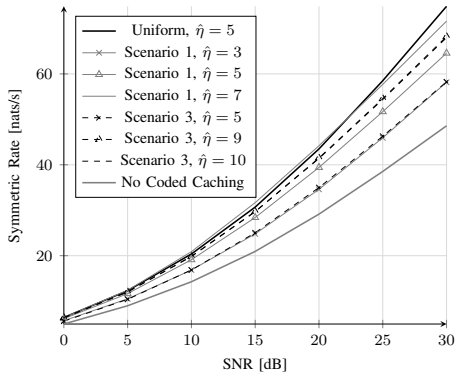

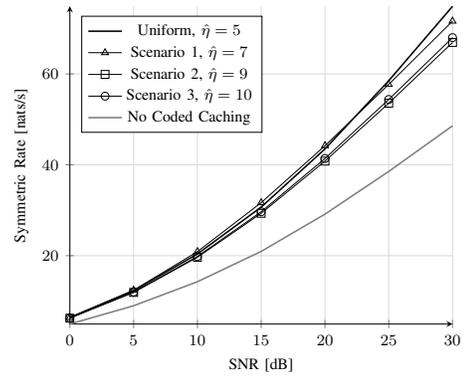
\begin{figure}[t]
    \centering
    \resizebox{0.7\columnwidth}{!}{%
    
    \begin{tikzpicture}

    \begin{axis}
    [
    axis lines = left,
    xlabel = \smaller {SNR [dB]},
    ylabel = \smaller {Symmetric Rate [nats/s]},
    ylabel near ticks,
    legend pos = north west,
    ticklabel style={font=\smaller},
    grid=both,
    major grid style={line width=.2pt,draw=gray!30},
    ]
    
    \addplot[black,thick]
    table[y=A5,x=SNR-dB]{Data/L12.tex};
    \addlegendentry{\smaller Uniform, $\hat{\eta} = 5$}
    
    \addplot[mark = triangle,black]
    table[y=B7,x=SNR-dB]{Data/L12.tex};
    \addlegendentry{\smaller Scenario 1, $\hat{\eta} = 7$}
    
    \addplot[mark = square,black]
    table[y=C9,x=SNR-dB]{Data/L12.tex};
    \addlegendentry{\smaller Scenario 2, $\hat{\eta} = 9$}
    
    \addplot[mark = o,black]
    table[y=D10,x=SNR-dB]{Data/L12.tex};
    \addlegendentry{\smaller Scenario 3, $\hat{\eta} = 10$}
    
    \addplot[gray,thick]
    table[y=NoCC,x=SNR-dB]{Data/L12.tex};
    \addlegendentry{\smaller No Coded Caching}

    \end{axis}

    \end{tikzpicture}
    }
    \caption{Performance comparison for $\hat{\eta} = \max \eta_p$, scenarios 1, 2, and 3}
    \label{fig:L12-ScenarioBCD}
\end{figure}

Simulation results for the proposed scheme are provided in Figures~\ref{fig:L12-ScenarioB} and~\ref{fig:L12-ScenarioBCD}. We have done simulations for a \ac{MISO} network with 50 single-antenna users, where the cache ratio at each user is $\gamma = 0.1$ (i.e., $P=10$), and the spatial \ac{DoF} is set to $\alpha = 10$. The number of transmitting antennas at the server can be any number larger than $\alpha$; it is assumed to be 12 here. For the distribution of $\eta_p$, we consider uniform distribution (i.e., $\eta_p = 5$ for $p \in [10]$) as well as three other scenarios, where the non-uniformity of $\eta_p$ values increase from scenario 1 to 3. The exact values of $\eta_p$ for these scenarios are shown in Table~\ref{tab:sim_scenarios}. In all simulations, optimized beamformers are used and both CC and unicast delivery phases are implemented. For a better comparison, in the figures we have also included the rate curve for the case all the required data is delivered through the unicast delivery phase (i.e., there is no coded caching gain).

\begin{table}[t]
    \centering
    \begin{tabular}{|c||c|c|}
        \hline
         Scenario & $\eta_p$ values & Standard deviation \\
         \hline
         \hline
         1 & 5,4,5,5,4,3,6,6,5,7 & 1.15 \\
         \hline
         2 & 9,3,1,4,5,7,2,6,5,8 & 2.58 \\
         \hline
         3 & 8,3,8,0,4,10,7,4,0,6 & 3.40 \\
         \hline
    \end{tabular}
    \caption{Simulation scenarios with non-uniform profile lengths}
    \label{tab:sim_scenarios}
\end{table}

In Figure~\ref{fig:L12-ScenarioB}, the effect of the $\hat{\eta}$ parameter on the performance in scenarios 1, 3 is shown. In general, choosing a larger $\hat{\eta}$ value (up to $\max \eta_p$) improves the performance. This is because with increased $\hat{\eta}$, a better coded caching gain, strong enough to cover for the \ac{DoF} loss of phantom users, is achieved. Of course, it should be noted that in scenario 3, the case $\hat{\eta} = 9$ provides slightly (about one percent) better rate compared with $\hat{\eta} = 10$. This is because as $\eta_p$ distribution becomes more non-uniform, by choosing $\hat{\eta} = \max \eta_p$ we need to add too many phantom users, increasing the chance of \ac{DoF} loss. Nevertheless, even for the very non-uniform case of scenario 3, this performance loss is negligible, and one can safely consider $\hat{\eta} = \max \eta_p$ in all cases. Of course, a more theoretical analysis, due for the extended version of this paper, would better clarify the probable performance loss of selecting $\hat{\eta} = \max \eta_p$.

Finally, in Figure~\ref{fig:L12-ScenarioBCD}, we compare the performance of the proposed scheme with the case of uniform $\eta_p$ distribution (i.e., the maximum possible coded caching gain), for scenarios 1, 2, and 3, when $\hat{\eta} = \max \eta_p$. As can be seen, the proposed scheme performs very well, with loss of less than ten percent for the considered network setup, compared with the uniform case. This is because with $\hat{\eta} = \max \eta_p$, we try to maximize the achievable global caching gain. Also, when \ac{DoF} falls due to phantom users, optimized beamformers enable a larger beamforming gain (as the size of the interference indicator set becomes smaller than $\alpha-1$), thus compensating for the minor performance loss. Interestingly, in the low-SNR regime and for scenario 1, the proposed scheme even outperforms the uniform case by a small margin. This is because the beamforming gain is more effective in this regime, as discussed in~\cite{tolli2017multi}.

\section{Conclusion and Future Work}
We introduced a new multi-antenna coded caching scheme for dynamic networks where the users are allowed to join and leave the network freely. Instead of dictating individual caching profiles for users, we associated them with a set of predefined caching profiles. Data delivery is then performed in two consecutive phases, with the goal of maximizing the achievable performance. The proposed scheme requires a small subpacketization, enables easy implementation of optimized beamformers, and its performance and robustness under dynamic conditions is shown through simulations. The current paper provides a proof-of-concept, with many theoretical analyses and optimizations due for future work. 

\bibliographystyle{IEEEtran}
\bibliography{references}

\end{document}